# Generation of femtosecond spin-polarized current pulses at Fe/MgO interface by quasi-static voltage

*Piotr Graczyk, Maria Pugaczowa-Michalska*

Institute of Molecular Physics, Polish Academy of Sciences, M. Smoluchowskiego 17, 60-179 Poznan, Poland

*Maciej Krawczyk*

Faculty of Physics, Adam Mickiewicz University, Poznań, Uniwersytetu Poznańskiego 2, Poznań 61-614, Poland

Abstract

The generation of short spin-current pulses is essential for fast spintronic devices. So far, spin current pulses are generated by femtosecond laser pulses which drive spins from a ferromagnetic metal layer. However, the need for miniaturization, simplicity and energy efficiency favour electric-field control of spintronic devices over optic or thermal control. Here, we combine *ab initio* calculations of electronic density of states at MgO/Fe interface with continuous model for charge transport to investigate the dynamics of the spin-dependent potential. We demonstrate that the voltage-driven instability of the electronic band structure due to the electronic resonant states at the Fe/MgO interface results in the generation of the femtosecond spin-polarized current pulse with the spin polarization up to $P = 700\%$ that propagates from the interface to the bulk. The dynamics of the current pulses driven by the Stoner instability depends neither on the dielectric relaxation time nor on the details of how the instability is achieved by changing the voltage, i.e. as long as the voltage changes are slow (quasi-static) with respect to the time determined by the spin diffusion constant, being of the order of fs.The presence of the instability may be detected by transport or magnetic measurements.

**INTRODUCTION**

Exchange interaction between electrons leads to the splitting of the *d* subbands and appearance of the spontaneous magnetization in the ferromagnetic transition metals like iron or nickel. The resultant asymmetry in the density of states for spin-up and spin-down bands at the Fermi level is the source of many spin-dependent effects, some of them already in the commercial use. Magnetic tunnel junctions[1] (that relies on the effect of spin-dependent tunneling) or spin valves[2] (that relies on the effect of spin-dependent conductivity) are the basis of the hard disk drives read-heads and magnetoresistive random-access memories (MRAM)[3].

The interaction between *d*- and *s*-band electrons results in the polarization of the conductive band *s*[4]. Therefore, the spin angular momentum may be transferred across or outside ferromagnetic medium as a spin(-polarized) current. It can be used to excite magnetization or move solitons via spin-transfer torque[5–9] or it can be transformed to the charge current by the inverse spin-Hall effect[10]. The femtosecond laser pulses[11–20] or spin pumping from the antiferromagnet[21] are used to generate subpicosecond spin-current pulses, which are regarded in spintronics as a state variable for signal transmission and processing at the ultrafast timescale or as a spintronic terahertz radiation emitters. However, optical lasers induce incoherent thermal effects, pose additional challenges in lowering the power consumption and device miniaturization. Here, we demonstrate the generation of the femtosecond spin-polarized current pulse from the quasi-static voltage taking advantage of the electronic resonant states at the Fe/MgO interface.

In Ref. 22 we showed that the spin-dependent potential at a dielectric-ferromagnetic metal interface contributes to the spin-polarized current in nanocapacitor containing ferromagnetic electrode when subjected to the ac voltage. It is assumed, like in most models concerning spin transport in metallic nanostructures[23,24], that the density of states, and therefore the conductivity of the ferromagnet, is a constant value, i.e. the electron kinetic energy does not change from the Fermi level. While it is good assumption for many problems of electron transport in metals, it may be not true if the significant accumulation of charge is present in the system, i.e. at the ferromagnet interface. In particular, as we discuss below, the energy-dependence of the density of states is crucial in transport of electrons between ordered interfaces[25,26]. Here, using energy-dependent density of states obtained from ab-initio calculations and the spin-dependent one-dimensional transport model, we show that the voltage applied to the MgO nanocapacitor with epitaxial Fe electrode leads to the Stoner instability at Fe/MgO interface. This instability occurs when the Fermi level exceeds the resonant interface electronic state. Importantly, this interfacial Stoner instability manifests itself in the spin-polarized current pulses, which can be detected with the charge current measurements, spin-Hall effect measurements, or other measurements technics sensitive to the spin or magnetization at interfaces.

## STONER MODEL AND THE SPIN-DEPENDENT POTENTIAL

The energy of the electron system in the ferromagnetic metal is the sum of the kinetic, exchange and electrostatic terms[27,28]:

$$E(\mu_\uparrow, \mu_\downarrow, \phi) = E_k + E_w + E_c = \int_0^{\mu_\uparrow} \mu \, \rho(\mu) d\mu + \int_0^{\mu_\downarrow} \mu \, \rho(\mu) d\mu - \frac{1}{2} eI \left( \int_0^{\mu_\uparrow} \rho(\mu) d\mu - \int_0^{\mu_\downarrow} \rho(\mu) d\mu \right)^2 - e \left( \int_0^{\mu_\uparrow} \rho(\mu) d\mu + \int_0^{\mu_\downarrow} \rho(\mu) d\mu \right) \phi, \quad (1)$$

where $\sigma = \uparrow, \downarrow$ is for spin-up and spin-down, respectively, $e$ is electron charge, $\mu_\sigma$ are chemical potentials, $\rho_\sigma$ are densities of states (DOS) that depend on $\mu_\sigma$, $I$ is Stoner parameter and $\phi$ is electric potential. With the specific number of electrons, it is possible, by calculation of the condition for the extremum of Eq. 1 to show that the bands fill up equally until the Stoner criterion $eI\rho > 1$ is reached. Then, the system becomes unstable and all the electrons fill up spin-up bands at the expense of the spin-down band filling. At this point the material becomes ferromagnetic. At some level of the band filling the Stoner criterion is no more satisfied and the spin-down bands start to fill up. Finally, the equilibrium state $E(\mu_\uparrow^0, \mu_\downarrow^0, 0)$ is reached.

With the external voltage applied and varying electric potential $d\phi$, we calculate the energy change as $dE = E(\mu_\uparrow^0 + d\mu_\uparrow, \mu_\downarrow^0 + d\mu_\downarrow, d\phi) - E(\mu_\uparrow^0, \mu_\downarrow^0, 0)$ and setting for convenience $\mu_\uparrow^0 = \mu_\downarrow^0 = 0$ we obtain electrochemical potentials as:

$$d\bar{\mu}_\sigma \equiv \frac{dE}{d\mu_\sigma} = d\mu_\sigma - ed\phi \mp eI(\rho_\uparrow(\mu_\uparrow)d\mu_\uparrow - \rho_\downarrow(\mu_\downarrow)d\mu_\downarrow), \quad (2)$$

where $\rho_\sigma(\mu_\sigma) = \rho(\mu_\sigma^0 + \mu_\sigma)$. After setting $d\bar{\mu}_\sigma = 0$ and solving for $d\mu_\sigma$ we obtain electron densities induced by potential $\phi$:

$$dN_\sigma = \frac{1 - 2eI\rho_{\sigma\prime}(\mu_{\sigma\prime})}{1 - eI\rho} e\rho_\sigma(\mu_\sigma)d\phi. \quad (3)$$

From Eq. (3) it comes that the induced electron density is increased for the band of the higher DOS and decreased for the band of the lower DOS in comparison to the densities obtained in the absence of the exchange term ($I = 0$). This effect is described by the difference in the electric potential felt by the spin-up and spin-down bands. From Eq. (2) we can define the local spin-dependent potential in the form:

$$\phi_\uparrow = \phi - \frac{I}{e}(n_\uparrow - n_\downarrow), \quad (4a)$$

$$\phi_\downarrow = \phi + \frac{I}{e}(n_\uparrow - n_\downarrow), \quad (4b)$$

where $n_\sigma = -e \int_0^{\mu_\sigma} \rho_\sigma(\mu) d\mu$ is an electron charge density.

The spin-dependent potential modifies the magnetization change at the interface of ferromagnetic metal and dielectric in the charge-mediated magnetoelectric effect[29–31]. It has influence on tunnelling in magnetic tunnel junctions[32], magnetocapacitance[33] and may contribute to the spin-polarized current[22]. However, in the examples above the density of states is assumed to be constant or parabolic, which is not accurate for epitaxial MgO/Fe heterostructure. It has been show, that at MgO/Fe interface the surface electronic states exist and they lead to the severe modulation of tunneling conductance[25,34], sign-reversal of the tunneling magnetoresistance[26], enhancement of the voltage-controlled magnetic anisotropy[35] and enhancement of the voltage-driven spin torque[36]. Therefore, we calculate DOS for MgO/Fe system from the first principles and we implement it as a function of chemical potential in the continuous model. With this we study the electronic surface states influence on the spin-polarized current generated in the system by the ac voltage in the non-tunnelling regime.

## THEORETICAL MODEL

### DFT CALCULATIONS

The density functional calculations were performed within the projector augmented wave (PAW) method[37] as implemented in Vienna ab initio Simulation Package (VASP)[38–40]. The generalized gradient approximation (GGA) in the Perdew-Burke-Ernzerhof (PBE) formulation[41] was assumed for description of exchange and correlations in the system. The Fe/MgO was modeled as a slab of 20 Å of iron atoms with 40 Å layers of MgO and vaccum region. We constructed a supercell model between Fe $bcc$ and MgO (NaCl structure). Obviously, in electronic structure calculations one is forced to use commensurable structures to model incommensurable systems[42]. The in-plane supercell was chosen such as to minimize the lattice mismatch between the iron and MgO. The [100] and [010] axes of the iron are rotated by 45° around [001] relative to corresponding axes of MgO. Vacuum region was set to avoid interactions between top and bottom atoms in the periodic slab. In making a commensurable supercell structure we adapt iron to the MgO lattice with lattice parameter 4.212Å at the first step. After that the cell shape and internal atomic coordinates were relaxed and the interface distance re-optimized until the residual forces on constituent atoms become smaller than 0.02 eV/Å. Supercell model used in further calculations was: Fe(001)~19Å/MgO(001)~40Å+vacuum~33Å. A kinetic energy cut-off of 520 eV and a total energy convergence threshold of $10^{-6}$ eV were used. Γ-centered k-grid mesh (7x7x1) was used for the Brillouin zone integration. The temperature blurring (at T= 300 K) of DOS was taken into account using approach described in Ref. [43].

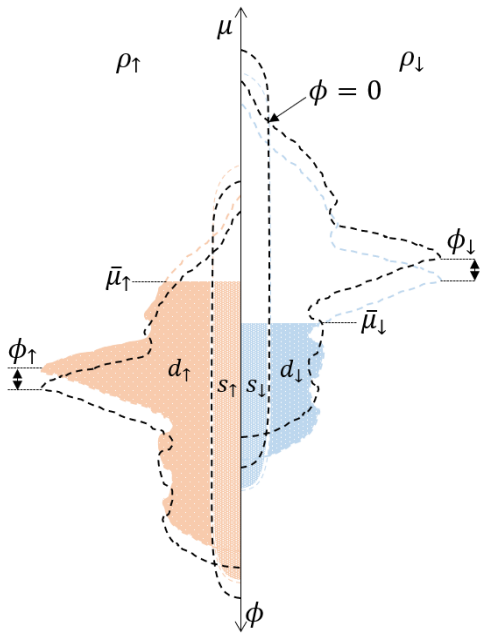

Fig. 1. Schematic plot of DOS in transition metals with the definition of chemical potentials in the continuous model and their change under accumulation of the screening charge, leading to the nonequilibrium state $\bar{\mu}_\uparrow - \bar{\mu}_\downarrow$. The dashed lines shows the initial position of the bands at $\phi = 0$.

CONTINUOUS MODEL

To simulate charge-current dynamics of the system we used one-dimensional two-current diffusive model in the s-d model approximation. The filling of the conductive bands is represented by the chemical potential $\mu_\sigma$ and its time-dependence is described by the continuity equation:

$$e\rho_\sigma \frac{\partial \mu_\sigma}{\partial t} = \frac{\partial J_\sigma}{\partial x} - \frac{e}{T_1} \frac{\rho_\sigma \rho_{\sigma'}}{(\rho_\sigma + \rho_{\sigma'})} (\bar{\mu}_\sigma - \bar{\mu}_{\sigma'}), \tag{5}$$

where $J_\sigma$ is charge density current, $T_1$ is spin-flip relaxation time and $\rho_\sigma = \rho_\sigma^d(\mu_\sigma) + \rho^s$ is density of states which consists of that which comes from bands $d$ and depends on chemical potential, and the constant $\rho^s$ from bands $s$. The absolute energy levels of the bands are represented by the electrochemical potential $\bar{\mu}_\sigma = \mu_\sigma - e\phi_\sigma$ where $\phi_\sigma$ is the spin-dependent potential (Fig. 1).

The current density $J_\sigma$ is described by the Ohm's law:

$$J_\sigma = e\rho^s D_\sigma \frac{\partial \bar{\mu}_\sigma}{\partial x} = e\rho^s D_\sigma \frac{\partial \mu_\sigma}{\partial x} - e^2 \rho^s D_\sigma \frac{\partial \phi_\sigma}{\partial x}, \tag{6}$$

where $D_\sigma$ is diffusion constant. We assume within s-d model that the current is carried by electrons in bands $s$ only. The charge densities are given by equations:

$$\frac{\partial n_\sigma}{\partial t} = -e\rho_\sigma \frac{\partial \mu_\sigma}{\partial t}, \tag{7}$$

and the electric potential obeys Gauss law:

$$-\epsilon_0 \frac{\partial^2 \phi}{\partial x^2} = n_\uparrow + n_\downarrow. \tag{8}$$

The dielectric layer is modelled by boundary conditions at its interfaces, i.e., continuity of the electric displacement:

$$-\epsilon_0 \frac{\partial \phi}{\partial x} = -\epsilon_0 \epsilon \frac{\phi_d - \phi_0}{d}. \tag{9}$$

The spin accumulation and spin current are defined as $s = -e\rho^s(\bar{\mu}_\uparrow - \bar{\mu}_\downarrow)$ and $J_s = e\rho^s \left( D_\uparrow \frac{\partial \bar{\mu}_\uparrow}{\partial x} - D_\downarrow \frac{\partial \bar{\mu}_\downarrow}{\partial x} \right)$, respectively. The magnetization change in Fe layer is $\Delta M =$

$\frac{\mu_B}{d}\int_0^{x_d}(\int_0^{\mu_\uparrow}\rho_\uparrow^d(\mu)d\mu - \int_0^{\mu_\downarrow}\rho_\downarrow^d(\mu)d\mu)dx$, where $\mu_B$ is Bohr magneton. The capacitance of the system is given by $C = \frac{A}{U}\int J dt$ where $J = J_\uparrow + J_\downarrow$ is the charge current, $U$ is applied voltage and $A$ is the capacitor surface. The polarization of the current is defined by $P = J_s/J \cdot 100\%$.

The simulated system is Cu/MgO/Fe/Cu multilayer. The MgO thickness is 4 nm to assure high density of the screening charges and, on the other hand, to avoid tunneling current[44,45]. The Fe thickness is 2 nm. Equations (4)–(9) are solved numerically by the finite-element method in COMSOL MULTIPHYSICS with sinusoidally time-varying voltage $U$ applied to the outer boundaries with amplitude $U_0 = 2$ V and frequency $f = 200$ MHz. This amplitude of voltage is below MgO dielectric breakdown value[46]. The relative permittivity for MgO is $\epsilon = 10$. The density of states of the s-bands is $\rho_s = 1.4 \cdot 10^{46}$ 1/J/m³ for Fe (taken from DFT simulations) and $\rho_s = 1 \cdot 10^{47}$ 1/J/m³ for Cu. The diffusion coefficients are $D_\uparrow = 0.0015$ m²/s and $D_\downarrow = 0.00021$ m²/s for Fe [47] and $D = 0.001$ m²/s for Cu [48]. The spin-flip relaxation times are $T_1 = 3.6$ ps for Fe and $T_1 = 250$ ps for Cu electrodes to assure spin relaxation lengths 60 nm and 500 nm, respectively. We assume rigid bands, i.e., that the DOS shape does not change with the applied voltage. We limited the presence of the Stoner instability to the point at the FM interface (i.e. $I = 0$ everywhere except the interface), which represents the first Fe atomic layer, since it is the unitary element which can fall into instability. The Stoner parameter is in the order of $I = \Delta/M \approx 10$ mV nm³ [28] and according to Ref. [49] it can be assumed to be a constant.

**RESULTS**

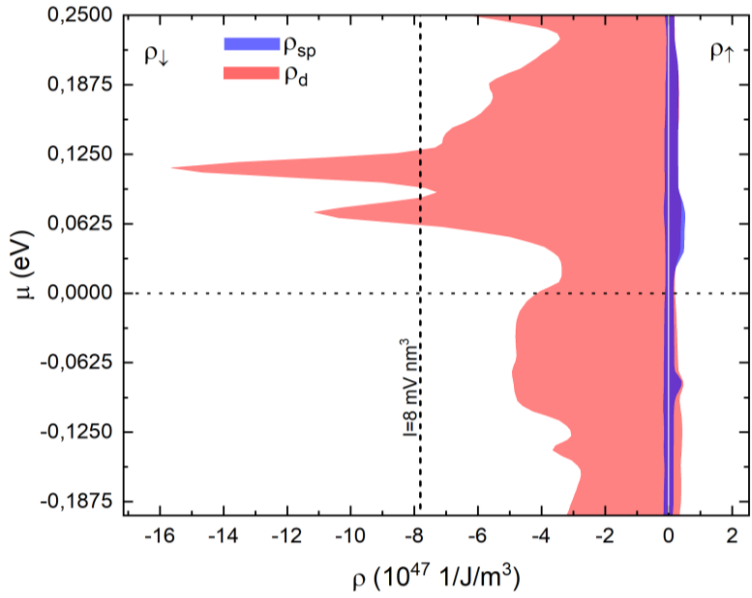

Fig. 2. Density of states obtained from DFT simulations at MgO/Fe interface divided to sp and d bands contributions. The vertical dashed line shows Stoner stability limit for $I = 8$ mV nm$^3$.

The DOS obtained from DFT simulations at Fe/MgO interface is shown in Fig. 2. The DOS for the band *d* is highly asymmetric with two resonant states at 0.0625 and 0.1100 eV for the minority band above the Fermi level.

The DOS from Fig. 2 has been used for the simulations in the continuous model. Fig. 3 shows the spin current changes in time under applied harmonic voltage assuming three different values of Stoner parameter, just to present its impact. The spin current value, $J_s$, is given in the Cu layer, 4 nm from the Fe/MgO interface. In the absence of the spin-dependent potential ($I = 0$) the spin current is cosine-like and its origin is mainly in the spin accumulation at Fe/Cu interface. Slight deviations from cosine are indications that DOS changes with applied voltage at MgO/Fe interface with changes of the band filling. For $I = 6$ mV nm$^3$ the amplitude of the spin current increases slightly as the contribution of the interface spin-dependent potential increases but the curve remains its continuous character. Unexpectedly, for $I = 8$ mV nm3 we observe sharp peak in the spin-current dependence on time. Interestingly, the peak is visible only for the time-range where the positive potential was applied on the ferromagnetic side (first half of the cycle), thus increasing number of electrons at MgO/Fe interface. Thus, observed peak of the spin current corresponds to the interface resonant states for $\rho_\downarrow^d$ electrons above Fermi level. The system becomes permanently unstable for $I > 8$ mV/nm$^3$ as the density of states significantly increases above Stoner criterion again for $\mu > 0.2$ eV.

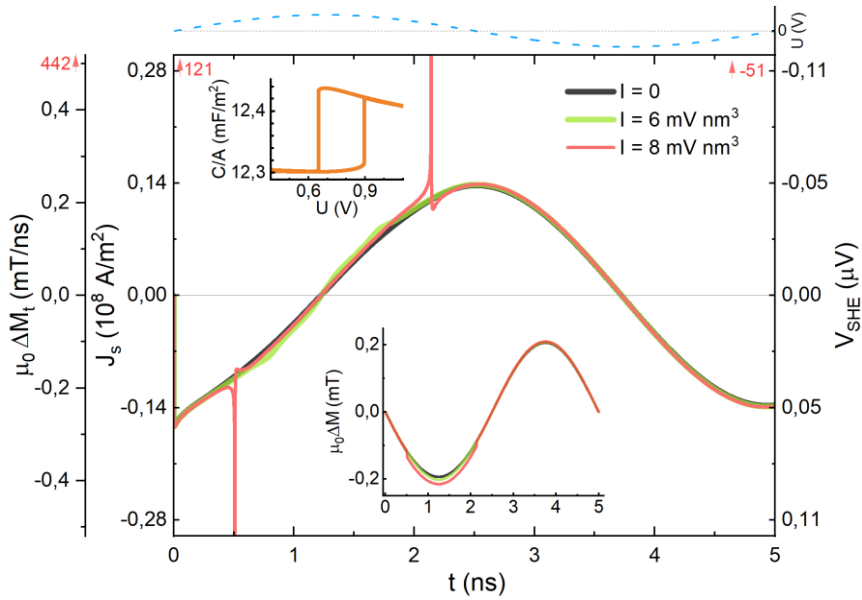

Fig. 3 Spin current $J_s$ in Cu in function of time for three values of Stoner parameter $I$ under time-varying voltage $U$ (top panel). The insets show the capacitance of the system in dependence on applied voltage $U$ (top left) and the time-dependence of magnetization change in Fe (bottom), while additional left axis shows its time derivative. The red labels at each vertical axis indicate maximum values of the peak, which exceed over 2 orders of magnitude the scales in the figure.

Together with the spin-current peak, the peak in the charge current is observed, which is indicated by the jump of the capacitance shown in the top inset in Fig. 3. Thus, the presence of the effect of Stoner instability should be detectable by the pure electrical measurements of the charge current. The hysteretic behaviour of the capacitance is discussed in the next paragraphs.

Fig. 4 shows the space- and time evolution of the spin-polarized pulse generated at the Fe/MgO interface. The amplitude of the spin current decays quickly within about 2 nm from the interface, but the long tail of the pulse spreads out to the several nanometers. The temporal width of the pulse is determined mainly by the diffusion constant $D$. If we assume the solution of the diffusion equation for spin accumulation without relaxation term $\partial s/\partial t = D\partial^2 s/\partial x^2$ as

$$s = \frac{1}{\sqrt{4\pi Dt}} e^{-\frac{x^2}{4Dt}}$$

then, calculating $J_s = -\partial s/\partial x$, taking half of the maximum value of $J_s$, we obtain the expression for the time that define full width at half of the maximum (FWHM) of the pulse:

$$\frac{3\sqrt{6}}{x^2} e^{-\frac{3}{2}} - x(Dt_{1,2})^{-\frac{3}{2}} e^{-\frac{x^2}{4Dt_{1,2}}} = 0 \ .$$

For $D = 4*10^{-4}$ m²/s and $x = 1$ nm the value of FWHM is $\Delta t = t_2 - t_1 \approx 1$ fs.

In the simulations, the temporal width of the pulse (left inset in Fig. 4) is equal to 9 fs at the distance of 1 nm from the interface and it increases quadratically with the distance from the interface, i.e., $\tau \propto x^2/D$. The polarization of the current (right inset in Fig. 3) is above the value of the polarization induced by the spin-dependent conductivity at the Fe/Cu interface ($P = -60\%$) to the distance up to 12 nm and it reaches values between 300% and 700% inside Fe layer. The value of $P$ exceeds 100% since the signs of $J_\uparrow$ and $J_\downarrow$ currents, driven by Stoner instability, are opposite.

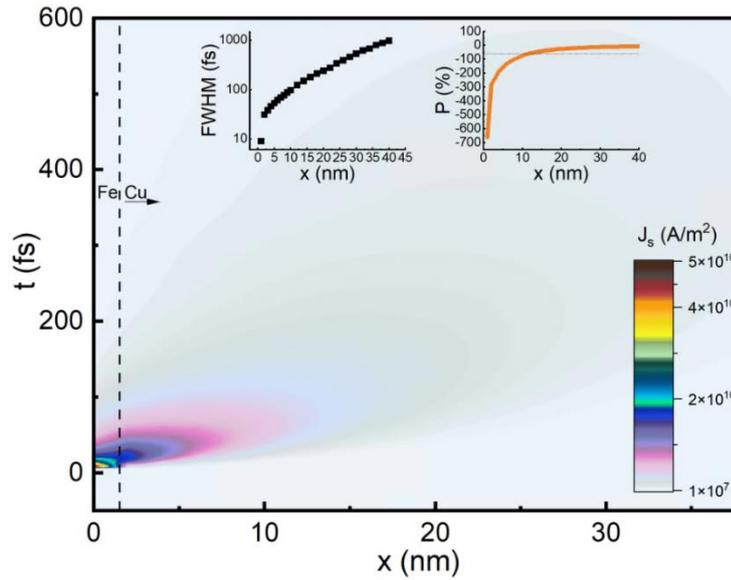

Fig. 4: The spin-polarized current pulse in dependence on time and distance from the Fe/MgO interface. The insets show the width of the pulse (left) and the polarization of the pulse (right) in dependence on the distance from the Fe/MgO interface with the polarization value generated by the spin-dependent conductivity at the Fe/Cu interface indicated by dashed line.

The relative positions of the peaks in time at 0.51 ns and at 2.14 ns in Fig. 3 does not correspond to the same value of the applied voltage, which is related to the fact that the additional electric energy is required to overcome energy barrier related to the Stoner instability and to recover the system to the initial state. This property is manifested by the hysteresis of the capacitance of the system shown in the inset of Fig. 3. Increasing the voltage from the zero value, the instability occurs spontaneously as soon as Stoner criterion at $U = 0.9$ V is reached from the bottom of the electronic resonance peak. However, when the voltage is decreased from the maximal value at $t = 1.25$ ns and the electrochemical potential $\bar{\mu}_\downarrow$ reaches top of the resonance peak, it is not possible to pump out the electrons from the minority band as it means reaching Stoner criterion that favours filling up minority band. Therefore, the majority band is excessively pumped-out of the electrons up to the voltage of $U = 0.65$ V at which the energy barrier is overcomed and the resonance peak in the minority band is emptied.

The asymmetry of the DOS below and above Fermi level is visible also from the asymmetric magnetization change $\Delta M$ of the Fe layer with time, i.e., for positive and negative voltages (inset in Fig. 3). The absolute value of magnetization change reaches a less than one militesla. Because of that asymmetry, the time-averaged magnetization change does not cancel out completely and reaches $\langle \Delta M \rangle_t = -2$ A/m in Fe for the highest value of Stoner parameter. However, this value is much beyond the measurement possibilities even for high-sensitivity magnetometers[50]. On the other hand, time-resolved techniques (e.g. magnetooptical Kerr effect[51]) should be feasible to detect asymmetric magnetization changes in Fe. The additional left axis in Fig. 3 shows the time derivative of $\Delta M$, which has the same time-dependence as spin current $J_s$. This indicates, that Stoner instabilities may be identified as discontinuities in the time dependence of $\Delta M$.

The well-known method to detect spin current is to use the inverse spin Hall effect to transform out-of-plane spin current into in-plane charge current and to measure the voltage induced by the charge flow. For this purpose it is possible to introduce a thin V or Pt layer between Fe and Cu. We roughly estimate the voltage from[23]:

$$V_{SHE} = \frac{-\theta_{SHA} \Delta y}{2e^2 \rho_s D} J_s, \tag{10}$$

assuming capacitor width $\Delta y = 2$ $\mu$m and the spin-Hall angle of vanadium $\theta_{SHA} = 0.01$ [52,53]. We assume that the magnetization is in-plane in the Fe layer and the spin current flow is perpendicular-to-plane only. The resulting spin Hall voltage is given by the right axis in Fig. 3.

The resulting spin-Hall voltage amplitude is in $\mu$V regime which is within today's measurement possibilities [54]. One can also consider replacement of vanadium electrode with platinum to increase $V_{SHE}$ as it has one order of magnitude higher $\theta_{SHA}$, if the growth of epitaxial Fe on Pt is possible. To obtain more quantitative estimation, one should take into account also other effects that deteriorate SHE signals, e.g., current shunting or spin memory loss.

Another technique for the experimental verification of the generation of the spin accumulation at MgO/Fe interface by time-dependent electric field would be the excitation of the sample with a THz radiation of proper polarization and measurement of the spin accumulation with the time-resolved

Kerr effect[51] or photoemission spectroscopy[55]. As the timescales in that experiments are orders of magnitude smaller than in presented simulations, one can expect significant enhancement of the effect.

An alternative way to generate spin-polarized current pulses by the abrupt capacitance change would be to make use of the polarization switching of the ferroelectric material in contact with Fe (e.g. $BaTiO_3$ [56–58]). However, the spin-polarized pulse width would be then inherently connected to the dielectric relaxation time and the rate of the polarization switching. It is not the case for the pulse generated by the Stoner instability. By introducing the term for the finite dielectric response to Eq. (9) in the form $-\epsilon_0 \tau_p \partial^2 V/\partial x \partial t$ with the typical dielectric relaxation time $\tau_p = 100$ ps we observe the shift of the position of the pulse in time, but with no change in its width. Since the pulse is driven by Stoner instability, also the frequency of the applied voltage does not play a role here (up to ~PHz range), in particular, it may be arbitrarily slow (quasi-static), while its amplitude must be just enough to reach the point of the instability.

We modelled the generation of the spin-polarized current at the MgO/Fe interface which is the result of the dynamic modification of the exchange splitting of the electronic bands via ac electric field. The strong energy-dependence of the $d$-band DOS at the Fermi level and interface resonant states lead to the sharp peaks in the spin-polarized current. This peaks can be directly measured as the charge current peaks or voltage peaks via the inverse spin Hall effect. The spin-polarized current peaks have a width in the femtosecond range and are highly spin-polarized. Importantly, the width of the pulse depends neither on the frequency of the applied voltage nor on the dielectric relaxation time. The Stoner instability at the Fe/MgO interface results in the hysteresis of the capacitance of the system. Our simulations indicate the possibility of generation of ultrashort spin-polarized current pulses from the electric signal operating in orders of magnitude longer timescale, which is of great interest for the applications in magnonics and spintronics.

The study has received financial support from the National Science Centre of Poland under grant 2018/28/C/ST3/00052.